\documentclass[superscriptaddress,showpacs,aps,twocolumn,floatfix,prl,longbibliography]{revtex4-1}
\usepackage{bm,amsmath,amssymb}
\usepackage{amssymb}
\usepackage{amsmath}
\usepackage{mathrsfs}
\usepackage{graphicx,bm}
\usepackage{verbatim}
\usepackage[dvipsnames]{xcolor}
\usepackage{soul}
\usepackage{ bbold }
\usepackage{xspace}
\usepackage{tikz}
\usetikzlibrary{snakes, arrows, backgrounds, calc, fadings, decorations.pathreplacing, shadings, through, intersections, angles, shapes.geometric}
\usepackage{pgfmath}
\usepackage{pgfplots}
\pgfplotsset{compat = newest}
\usepackage[T1]{fontenc}
\usepackage[utf8]{inputenc}

\usepackage[colorlinks]{hyperref}
\AtBeginDocument{%
 \hypersetup{ 
 linkcolor=blue,
 filecolor=magenta, 
 urlcolor=blue,
 citecolor=blue,
 colorlinks=true,
 }
}

\newcommand{\p}	{\partial}

\newcommand{\C}	{\mathbb{C}}
\newcommand{\Z}	{\mathbb{Z}}
\newcommand{\N}	{\mathbb{N}}

\newcommand{\cB}{\mathcal{B}}

\newcommand{\cD}{\mathcal{D}}

\newcommand{\cH}{\mathcal{H}}

\newcommand{\cO}{\mathcal{O}}

\newcommand{\cR}{\mathcal{R}}

\newcommand{\cV}{\mathcal{V}}

\newcommand{\cZ}{\mathcal{Z}}

\DeclareMathOperator{\Tr}{Tr}
\DeclareMathOperator{\tr}{\tr}

\renewcommand{\Tr}	{\mathrm{Tr}}
\renewcommand{\tr}	{\mathrm{tr}}

\newcommand{\bra}[1]	{\langle{#1}\vert}
\newcommand{\ket}[1]	{\vert{#1}\rangle}

\newcommand{\bh}    {\bar{h}}

\newcommand{\bq}    {\bar{q}}

\newcommand{\bT}    {\bar{T}}
\newcommand{\bi}    {\bar{\iota}}

\newcommand{\tq}    {\tilde{q}}

\newcommand{\ketbra}[2]	{\ket{#1}\bra{#2}}

\newcommand\id		{\mathbf{1}}

\newcommand{\vir}   {\mathsf{Vir}}

\newcommand{\bket}[1]	{\lVert{#1}\rangle\!\rangle}
\newcommand{\iket}[1]	{\lvert{#1}\rangle\!\rangle}

\newcommand{\gf}        {\mathsf{g}}

\newcommand{\cc}        {\mathsf{c}}
\newcommand{\cce}       {\cc_\mathsf{eff}}
\newcommand{\bcc}       {\bar{\cc}}
\newcommand{\prob}      {\mathsf{p}}
\newcommand{\probab}    {\prob^{\alpha\beta}}

\newcommand{\niab}      {\mathsf{n}^i_{\alpha\beta}}
\newcommand{\nab}       {\mathsf{n}_{\alpha\beta}}

\newcommand{\spec}      {\sigma}
\newcommand{\czab}      {\cZ^{\alpha\beta}}
\newcommand{\Sab}       {S^{\alpha\beta}}
\newcommand{\sab}       {s^{\alpha\beta}}
\newcommand{\Sfluct}    {S_{\textsf{f}}}
\newcommand{\Sconfig}   {S_{\textsf{c}}}
\newcommand{\Sfluctn}   {S_{\textsf{f},n}}
\newcommand{\Sconfign}  {S_{\textsf{c},n}}
\newcommand{\Suni}[1]   {S^{\verm}_{#1}}
\newcommand{\Sbdy}      {S_{\mathsf{bdy}}}
\newcommand{\Zab}       {Z_{\alpha\beta}}
\newcommand{\modS}      {\mathsf{S}}
\newcommand{\qdim}      {\cD}
\newcommand{\Stop}      {S_{\mathsf{top}}}
\newcommand{\ab}        {\alpha\beta}
\newcommand{\nul}       {\nu}
\newcommand{\fus}       {N}

\newcommand{\fusab}     {\fus_{\alpha\beta}}
\newcommand{\sspe}      {s^{\ab}}
\newcommand{\verm}      {\mathsf{Verma}}
\newcommand{\vac}       {\id}
\newcommand{\Zbulk}     {Z_{\text{bulk}}}

\newcommand\quotes[1]  {``{#1}''}

\newcommand\figref[1]	{figure~\ref{#1}\xspace}

\begin{document}

\title{Entanglement Resolution with Respect to Conformal Symmetry}

\author{Christian Northe}
\thanks{northe@post.bgu.ac.il}
\affiliation{Department of Physics, Ben-Gurion University of the Negev,
David Ben Gurion Boulevard 1, Be'er Sheva 84105, Israel}


\begin{abstract}
Entanglement is resolved in conformal field theory (CFT) with respect to conformal families to all orders in the UV cutoff. To leading order, symmetry-resolved entanglement is connected to the quantum dimension of a conformal family, while to all orders it depends on null vectors. Criteria for equipartition between sectors are provided in both cases. This analysis exhausts all unitary conformal families. Furthermore, topological entanglement entropy is shown to symmetry-resolve the Affleck-Ludwig boundary entropy. Configuration and fluctuation entropy are analyzed on grounds of conformal symmetry.
\end{abstract}

\maketitle
\textit{1. Introduction}.--- 
Since its discovery in 1935 \cite{Einstein:1935rr}, entanglement has been at the core of quantum theory \cite{schroedinger35}. In modern days, it is advancing our understanding of physics on many frontiers including phases of matter and quantum information \cite{QinfoQmatter} or gravity \cite{Rangamani:2016dms}, to name a few. It is thus important to distill the fineprints of entanglement. One recently successful route revolves around symmetries.
%
They organize the entanglement spectrum into various charge sectors, permitting to investigate how these contribute to entanglement \cite{Caputa:2013eka, Belin:2013uta, Goldstein:2017bua}. 
This so-called symmetry resolution of entanglement (SRE) has been applied so far to global internal symmetries in quantum field theories (QFT) and finite-dimensional systems \cite{Belin:2013uta, Goldstein:2017bua,Xavier:2018kqb,Capizzi:2020jed,Calabrese:2021wvi,Murciano:2020vgh,Horvath:2020vzs,Chen:2021nma,Chen:2021pls,Zhao:2020qmn,Weisenberger:2021eby,Zhao:2022wnp,Baiguera:2022sao,DiGiulio:2022jjd,Bonsignori:2020laa,Cornfeld:2018wbg,Bonsignori:2019naz,Murciano:2019wdl,Fraenkel:2021ijv,Tan:2019axb,Murciano:2022lsw,Azses:2021wav,Azses:2022nfl}.
Reassuringly, some results are already finding experimental realization \cite{Lukin19,Azses:2020tdz,Neven:2021igr,Vitale:2021lds}.

For abelian groups, entanglement is inspected with regards to fixed charge eigenvalues. Surprisingly, in all studied systems each charge sector contributes equally to entanglement to leading order in a UV cutoff $\epsilon$. This \textit{equipartition of entanglement} \cite{Xavier:2018kqb} is usually broken at $\cO(\epsilon)$. Turning to non-abelian groups $G$, the focus is shifted to organizing the entanglement spectrum into representations of $G$ \cite{Calabrese:2021wvi, Milekhin:2021lmq}. These are also equipartitioned to leading order, however only up to $\cO(1)$. Overall, equipartition is a ubiquitous feature, yet its origin remains somewhat elusive. For $U(1)$ CFTs it was recently associated with the Fock space structure \cite{DiGiulio:2022jjd}. 

In this letter, entanglement is resolved with respect to conformal symmetry in $(1+1)d$, which is a pillar of theoretical physics with numerous applications including critical phenomena \cite{Belavin:1984vu,Cardy:1996xt,DiFrancesco:1997nk,Cho:2016kcc}, strongly interacting systems \cite{Gogolin:2004rp, Senechal:1999us}, topological phases of matter \cite{Cho:2016xjw, PhysRevLett.101.010504, PhysRevLett.108.196402, Han:2019kyq}, non-equilibrium physics \cite{Gawedzki:2017woc,Bernard:2016nci} and through the AdS/CFT correspondence \cite{Maldacena:1997re}, it plays a key role even in gravity \cite{Rangamani:2016dms,Witten:2021nzp, Milekhin:2021lmq}.

An obvious choice, in line with the lore on SRE, is to resolve with respect to $L_0$ eigenspaces, corresponding to states of equal energy. In this paper, the entanglement spectrum is resolved instead with respect to irreducible representations of the Virasoro algebra $\vir$, i.e. conformal families, as this allows the full power of conformal symmetry to come to bear.
Indeed, this \emph{Virasoro resolution} is shown below to harbor remarkably rich physics. Two novelties, compared with conventional resolution, can be pointed out right away however.
First, conformal symmetry is a spacetime symmetry. Hence it may contain global and \emph{local} transformations \footnote{The reader might be tempted to argue that internal symmetries can also be local, i.e. be gauge symmetries. However, they are redundancies of the description rather than actual symmetries.}; $\vir$ indeed contains an infinity of the latter. Second, in contrast to conventional non-abelian SRE, the investigated representations are infinite-dimensional. This letter shows that such infinities present no hurdles for SRE. In fact, as so often in CFT, they are virtues.

Virasoro resolution is promising on many frontiers. For instance, it can indicate which families, or anyons, have the most relevance for ground state entanglement in gapped and topological phases of matter adjacent to critical points \cite{Cho:2016kcc}. By universality, these results extend to a plethora of systems. Turning to gravity, as shown below, the entanglement stored by the vacuum family is distinguished while all other families are equipartitioned. 

Estabilishing Virasoro resolution and elaborating its details is the subject of this letter. At $\cO(\epsilon^0)$, it is shown to be controlled by the topological entanglement entropy (TEE) \cite{Kitaev:2005dm, PhysRevLett.96.110405}. The requirements for equipartition between two conformal families are derived and exemplified in Virasoro minimal models. 
Remarkably, Virasoro resolution can be pushed to all orders in $\epsilon$, permitting an in-depth analysis of equipartition between two families, which explains its origin and even how it can be manufactured.
Furthermore, the TEE is shown to Virasoro-resolve the Affleck-Ludwig boundary entropy of the entanglement spectrum, thereby establishing the former as building blocks of the latter. Finally, complete equipartition of the entanglement spectrum is analyzed. Elaborate calculations are found in the supplemental material (SM) \cite{SM}. It is emphasized that Virasoro resolution is performed on general grounds and applies to entire classes of CFTs. Particular models are only drawn in as examples. 


\begin{figure}
 \includegraphics[scale=.38]{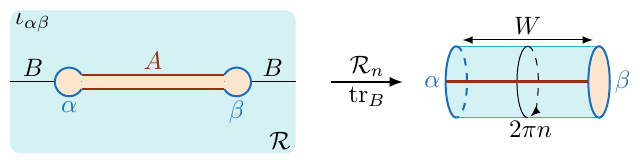}
 \caption{The Factorization $\iota_{\ab}$ imposes disks with boundary conditions $\alpha,\beta$, thereby placing the system on the manifold $\cR$. Replicating this to $\cR_n$, tracing over $\cH_{B,\ab}$ and a subsequent conformal transformation provides an annulus of width $W$ and circumference $2\pi n$.}
 \label{figAnnulus}
\end{figure}

\textit{2. Entanglement in QFT}.---
Entanglement requires a factorization into a product Hilbert space $\cH=\cH_A\otimes\cH_B$, which is usually associated with spatial domains $A$ and $B$. 
To implement this in QFT \cite{Ohmori_2015, Brustein:2010ms}, boundary conditions need to be assigned on the fields at the boundaries $\p A$ and $\p B$.
Specifying to $(1+1)d$ and a single entangling interval $A$, this is achieved by excizing the two entangling points comprising $\p A$ with two small disks of radius $\epsilon\ll1$ and imposing boundary conditions $\alpha$ and $\beta$ thereon \cite{Ohmori_2015}, see \figref{figAnnulus}. This manifold is called $\cR$ in the following. Formally, this is accomplished by a factorization mapping on Hilbert space
\begin{equation}\label{iota_def}
 \iota_{\alpha\beta}: \cH \to \cH_{A,\alpha\beta}\otimes\cH_{B,\alpha\beta},
 \qquad
 \iota_{\ab}:\,\ket{\psi}\mapsto\iota\ket{\psi},
\end{equation}
for $\ket{\psi}\in\cH$. In this way, the subfactor Hilbert spaces depend on boundary conditions in a QFT, as is confirmed numerically \cite{Lauchli:2013jga,Roy_2020,Hu:2020suv}. Using QFT as continuum limit of lattice systems, the boundary conditions $\alpha,\beta$ descend from boundary conditions imposed at the interval ends $\p A$ at the level of the lattice, as exemplified below.

A density matrix is reduced to $\cH_{A,\alpha\beta}$ by tracing over $B$, $\rho_A:=\Tr_{\cH_{B,\alpha\beta}}\left[\iota\ket{\psi}\bra{\psi}\iota^\dagger\right]$. 
Entanglement between degrees of freedom in $A$ and $B$ is quantified for pure states by the R\'enyi entropies 
\begin{equation}
 S_n^{\alpha\beta}
 =
 (1-n)^{-1}\log\tr\rho_A^n
 =
 (1-n)^{-1}\log\left[\frac{Z^{\alpha\beta}_n}{(Z_1^{\alpha\beta})^n}\right] ,
\end{equation}
where the superscripts associate it with the factorization $\iota_{\alpha\beta}$, $\tr$ denotes the trace over $\cH_{A,\alpha\beta}$ and $Z_n^{\alpha\beta}$ is the partition function of the QFT on an $n$-fold replicated manifold $\cR_n$ of $\cR$.

\textit{3. Conformal Factorizations}.---
The focus of this note lies on $(1+1)d$ CFTs. They have symmetry algebra $\vir\times\vir$, generated by the energy-momentum tensors $T,\, \bT$ and a field content captured by a bulk partition function $\Zbulk(q,\tq)=\sum_{i,\bi}M_{i\bi}\chi_i(q)\chi_{\bi}(\bq)$.
It has conformal characters $\chi_i(q)=\tr_{\cH_i}\left[q^{L_0-\cc/24}\right]$, where $\cH_i\equiv \cH(\cc,h_i)$ is an irreducible Virasoro module at conformal weight $h_{i}$ and central charge $\cc$. 

In this text, the entanglement structure of the vacuum in $\Zbulk(q,\bq)$, denoted $\ket{h_i=0,h_{\bi}=0}$ with density matrix $\rho=\ketbra{0,0}{0,0}$, is investigated. Furthermore, factorizations $\iota_{\ab}$ given by conformal boundary conditions $\alpha,\beta$.
They require $\bT=T$ at the disks and $\bcc=\cc$. Such $\iota_{\ab}$ are referred to as \emph{conformal factorizations}. Their relevance for entanglement spectra of gapped phases renders them important even away from critical points \cite{Cho:2016kcc,Cho:2016xjw}. 

It is useful to restrict to the setups studied in \cite{Cardy:2016fqc} for which $\cR$ can be conformally mapped to an annulus of width $W$ and height $2\pi$, which enter computations via $ q =
 e^{-2\pi^2/W}, \, \tq = e^{-2W}$. 
E.g., for a density matrix $\rho$ at zero temperature and on the infinite real line, $W=2\log(\text{length}(A)/\epsilon)$. $Z^{\ab}_n$ is then calculated on an annulus of height $2\pi n$ and width $W$, see \figref{figAnnulus}.
In this case, the reduced density matrix is
\begin{equation}\label{densityMatrix}
 \rho_A=\frac{q^{L_0-c/24}}{Z_{\alpha\beta}(q)}\,,
 \quad
 Z_{\alpha\beta}(q)=\sum_{i\in \spec}\niab\,\chi_i(q)
\end{equation}
where $\Zab(q)=\tr q^{L_0-c/24}$ quantifies the entanglement spectrum $\cH_{A,\alpha\beta}
=\bigoplus_{i\in \spec}\cH_i^{\oplus \niab}$, with multiplicities $\niab\in\N$,
imposed by \eqref{iota_def}. The normalization secures $\tr \rho_A=1$. If the set of all conformal families at central charge $\cc$ is denoted $I$, then $\spec\subset I$ denote the conformal families occuring in the entanglement spectrum. The ground state $\ket{\Omega}$ in the entanglement spectrum is the primary of lowest conformal dimension. For non-unitary theories it can be negative, $h_\Omega<0$ \cite{DiFrancesco:1997nk}. 

To illustrate this machinery, consider the critical Ising chain $H=\sum_{j\in\Z}(\sigma_j^z\sigma_{j+1}^z+\sigma_j^x)$, which corresponds to the Ising CFT in the continuum. The CFT Hilbert space $\cH$ in \eqref{iota_def} contains three primaries, $\spec=\{\id, \epsilon,\tilde{\sigma}\}$, which in turn label the boundary conditions, $\alpha\in\spec$. When choosing a subregion $A$ in the chain, the spins at the interval ends $\p A$ are typically left free. This descends to the CFT boundary conditions $\alpha=\beta=\tilde{\sigma}$, i.e. it pertains to a factorization $\iota_{\tilde{\sigma}\tilde{\sigma}}$ with entanglement spectrum $Z_{\tilde{\sigma}\tilde{\sigma}}=\chi_\id+\chi_\epsilon$ \cite{Ohmori_2015}. On the other hand, the spins in the chain can also have fixed boundary conditions at $\p A$. Equal orientation of these two spins leads to the CFT entanglement spectrum $Z_{\id\id}=\chi_0$, while opposite orientation descends in the CFT to $Z_{\id\epsilon}=\chi_\epsilon$. Lastly, fixing the spin at one end of the chain and leaving it free at the other leads to $Z_{\id\tilde{\sigma}}=\chi_{\tilde{\sigma}}$ in the CFT \cite{Ohmori_2015}.

Returning to generality, this framework readily provides R\'enyi entropies for $\rho=\ketbra{0,0}{0,0}$ \cite{Ohmori_2015},
\begin{align}
 \Sab_n
 =
 \frac{1}{1-n}\log\left(\frac{Z_{\alpha\beta}(q^n)}{(Z_{\alpha\beta}(q))^n}\right)
 \approx 
 S^{\Omega}_n
 +
 \Sbdy^{\ab}
 +\dots
 \label{Renyi}
\end{align}
While, the first equality is valid for any CFT, the approximation is valid only for rational CFTs, for which the set of conformal families $I$ is finite. It applies in the limit $q\to1^-$ and $\tq\to0^+$, where the character of the ground state $\Omega$ dominates the entanglement spectrum. The boundary entropy, $\Sbdy^{\ab}=\log[\gf_\alpha\gf_\beta]$, furnished by the Affleck-Ludwig g-factors \cite{Affleck:1991tk}, is a measure of the (asymptotic) size of the entanglement spectrum associated with the factorization $\iota_{\alpha\beta}$. It cannot be absorbed for all $n$ by rescaling of $\epsilon$ and is thus physical \cite{Cardy:2016fqc}. The first summand is responsible for the well-known leading behavior of entanglement entropy \cite{Calabrese:2009qy, Ohmori_2015},  
\begin{equation}\label{SOmega}
 S^{\Omega}_n
 =
 (1-n)^{-1}\log\left[\frac{\chi_\Omega(\tq^{1/n})}{(\chi_\Omega(\tq))^n}\right]
 \approx
 \frac{\cce W}{12}\frac{n+1}{n}+\dots
\end{equation}
where $\cce=\cc-24h_{\Omega}$ and $W\to\infty$ as $\epsilon\to0$.

\textit{4. Virasoro resolution}.--- The remainder of this letter explains how the conformal families comprising the entanglement spectrum, i.e. the set $\spec$, contribute to entanglement in the state $\rho=\ketbra{0,0}{0,0}$.

As implicitly used in \eqref{densityMatrix} for conformal factorizations, $\Zab$ decomposes into irreducible Virasoro representations $i$. Because the reduced density matrix \eqref{densityMatrix} lies in the (global) conformal group, it decomposes into block-diagonal form
\begin{align}
 \rho_A
 =
 \bigoplus_{i\in\spec}\Pi_i\rho_A
 =
 \bigoplus_{i\in \spec}\probab_i\,\rho_A(i)
\end{align}
If $\{\ket{i,m}\}$ is a basis for $\cH_i$ labelled by $m\in\N_0$, then $\Pi_i=\sum_{m=0}^\infty\ketbra{m,i}{m,i}$ is a projector onto $\cH_i$. Each family $i$ has a density matrix block 
\begin{equation}\label{densityMatrixBlock}
 \rho_A(i)=(\niab\chi_i(q))^{-1}q^{L_0-\cc/24}
\end{equation}
with $L_0$ restricted to $\cH_i^{\oplus\niab}$, securing that $\tr\rho_A(i)=1$. Such blocks comprise the entanglement degrees of freedom in the family $i$, measured with probability
\begin{equation}\label{probabilities}
 \probab_i=\tr[\Pi_i\rho_A]=\niab\frac{\chi_i(q)}{Z_{\alpha\beta}(q)}
 \quad
 \text{ for }
 i\in\spec
\end{equation}
and zero otherwise. This uses \eqref{densityMatrix} and the reader is reminded that $\tr=\tr_{\cH_{A,\alpha\beta}}$. Clearly, $\sum_{i\in\spec}\probab_i=1$. 
The probabilities \eqref{probabilities} are the $n=1$ case of charge-projected partition functions, 
\begin{align}\label{ChargedZ}
 \czab_n(i)
 :=
 \tr[\Pi_i\rho_A^n]
 =
 \niab\frac{\chi_i(q^n)}{(\Zab(q))^n}
 \quad
 \text{ for }
 i\in\spec\,.
\end{align}
They give rise to the symmetry-resolved R\'enyi entropies (SRRE)
\begin{align}
\Sab_n(i)
&:=
\frac{\tr[\rho_A(i)^n]}{1-n}
=
\frac{1}{1-n}\log\left(\frac{\czab_n(i)}{(\czab_1(i))^n}\right)\notag\\
&\,=
\frac{1}{1-n}\log\left((\niab)^{1-n}\frac{\chi_i(q^n)}{(\chi_i(q))^n}\right)
\label{SRE}
%
\end{align}
This is the central result of this letter and, after two remarks, it is shown to bear remarkably rich physics.

1) One charge operator for Virasoro resolution is the entanglement Hamiltonian $K_A=\frac{\pi}{W}(L_0-\cc/24+const.)$ itself, though eigenvalues of $L_0$ are taken modulo integers. This identification associates all descendants generated by the entanglement $\vir$ algebra \cite{Hu:2020suv} with a primary $\ket{h_i}$. Hence, Virasoro sectors are easily read out from the entanglement spectrum by rewriting each energy $\xi$ as $\xi=h_i+n$ for $n\in\N_0$, thereby connecting any $\xi$ to one family $i\in\spec$. The central charge (operator) is a second charge operator. Together, the pair $(h_i,\cc)$ uniquely identifies a conformal family.

2) Charged moments are completely bypassed in this text by the powerful tools of representation theory customary in CFT. This approach was pioneered in \cite{DiGiulio:2022jjd} for the simple case of $U(1)$ symmetry and is further developed here to suit conformal symmetry. In particular, conformal characters have more structure than their $U(1)$ counterparts. This provides profound insight into \eqref{SRE} in what follows.

\textit{5. Asymptotic equipartition and TEE}.---
The SRRE \eqref{SRE} is now analyzed for rational CFTs with respect to $\vir$, meaning that $I$ is finite, and in the limit $q\to1^-$ ($\tq\to0^+$); this is referred to as \quotes{asymptotic} and corresponds to $\epsilon\to0^+$. To that end, introduce the modular S-matrix via $\chi_i(q)=\sum_{k\in I}(\modS^{-1})_{ik}\chi_k(\tq)\approx \modS_{i\Omega}\, \chi_{\Omega}(\tq)$ \cite{SM}.
The SRRE \eqref{SRE} is thus approximated by 
\begin{align}
\Sab_n(i)
\approx
S^\Omega_n+\Stop^{\ab}(i)+\dots
\label{SREasymp}
\end{align}
The appearance of $S^\Omega_{n}$ explains the origin of leading-order equipartition for \emph{all} rational CFTs. The second term is $\cO(\epsilon^0)$, cannot be absorbed for all $n$ by rescaling $\epsilon$, and is thus physical. It is the TEE \cite{Kitaev:2005dm, PhysRevLett.96.110405}, 
\begin{equation}\label{Stop}
 \Stop^{\ab}(i):=-\log(\qdim/(\niab\qdim_i))=\log(\niab\modS_{i\Omega})
\end{equation}
built from the quantum dimensions
$
\qdim_i
:=
\modS_{i\Omega}/\modS_{\vac\Omega}
$
and total quantum dimension 
$\qdim\equiv(\sum_{i\in I}\qdim_i^2)^{1/2}=(\modS_{\vac\Omega})^{-1}$
\footnote{The TEE is usually defined as $-\log(\qdim/\qdim_i)$. Since it counts the entropy coming from one type of anyon, or in this case conformal family,  it is natural to include the multiplicities $\niab$ into the definition \eqref{Stop}. This corroborates that the TEEs depend on the factorization $\iota_{\alpha\beta}$. For unitary theories $\Omega=\id$, so that the definition of $\qdim_i=\modS_{i\Omega}/\modS_{\vac\Omega}$ reduces to the standard definition in the literature $\qdim_i=\modS_{i\vac}/\modS_{\vac\vac}$. The definitions given in the text are natural extensions to non-unitary theories.}. The $\qdim_i$ measure the \quotes{asymptotic size} of the family $i$ in relation to the vacuum family $i=\vac$ \cite{SM}. 

This provides a simple rationale to determine if two conformal families $i$ and $j$ are asymptotically equipartitioned, i.e. $\Sab_n(i)=\Sab_n(j)$ at $\tq\to0^+, (\epsilon\to0^+)$. \emph{Asymptotic ij-equipartition} occurs if
\begin{equation}\label{AsympEP}
 \Stop^{\ab}(i)=\Stop^{\ab}(j)
 \text{ for }
 i,j\in\spec
 \text{ and }
 i\neq j\,.
\end{equation}
Since the multiplicities $\niab$ are integer and the $\qdim_i$ are usually not, \eqref{AsympEP} entails two requirements. Families of equal asymptotic size, $\qdim_i=\qdim_j$, and equal \quotes{weight} in the entanglement spectrum $\Zab$,  $\niab=\nab^j$, harbor the same amount of information at leading order in the UV cutoff $\epsilon$ in the factorization $\iota_{\ab}$. This is emphasized by noting that \eqref{AsympEP} can be rephrased as $\probab_i=\probab_j$ with asymptotic probabilites, as derived from \eqref{probabilities},
\begin{equation}\label{probabilitesAsymp}
 \probab_i
 \approx
 \frac{\niab\modS_{i\Omega}}{\gf_\alpha\gf_\beta}
 =
 \frac{\niab\qdim_i}{\sum_{j\in\spec}\nab^j\qdim_j}
\end{equation}

A large class of solutions to \eqref{AsympEP} for Virasoro minimal models is provided in the SM \cite{SM}. 
The simplest example thereof arises in the Ising CFT with its three primaries $\id$, $\varepsilon$, $\tilde{\sigma}$ and $\modS_{\id\id}=\modS_{\id\varepsilon}=1/2$ \cite{DiFrancesco:1997nk}.
It possesses a Cardy boundary condition labelled by $\tilde{\sigma}$ providing an entanglement spectrum $Z_{\tilde{\sigma}\tilde{\sigma}}=\chi_\id+\chi_\varepsilon$ \cite{Cardy:1989ir} pertaining to a factorization $\iota_{\tilde{\sigma}\tilde{\sigma}}$.
Hence $\mathsf{n}^\id_{\tilde{\sigma}\tilde{\sigma}}=\mathsf{n}^\varepsilon_{\tilde{\sigma}\tilde{\sigma}}=1$ which secures asymptotic $\id\varepsilon$-equipartition. As explained above, this setup corresponds to an Ising chain with free boundary conditions imposed at the interval ends.

\textit{6. Exact equipartition}.---
In order to analyze SRREs to all orders, and beyond the regime of rational CFT, it is crucial to recall the relation between irreducible Virasoro modules $\cH_i$ and null vectors. The former is obtained after appropriately quotienting out all null vectors in the Verma module $\cV_i$ for conformal weight $h_i$\,; details are reviewed in the SM \cite{SM}. The character for $\cH_i$ reflects this null vector structure
\begin{equation}\label{nullStructure}
 \chi_i(q)=\frac{q^{\frac{1-\cc}{24}}}{\eta(q)}\nul_i(q),
 \quad
 \eta(q)=q^{1/24}\prod_{k=1}^\infty(1-q^k)
\end{equation}
via a function $\nul_i(q)$. For instance, when $\cH_i=\cV_i$ is a Verma module, i.e. it has no singular vectors, then $\nul_i=q^{h_i}$, and when $\cV_i$ carries a single null vector at level $N$, $\nul_i(q)=q^{h_i}(1-q^N)$. As usual, the Dedekind eta $\eta(q)$ keeps track of descendants.

Plugging this structure into the SRRE \eqref{SRE}, it becomes
\begin{equation}\label{SRREnull}
 \Sab_n(i)
 = 
 \frac{1}{1-n}\log\left[\frac{\nul_i(q^n)}{\nul_i(q)^n}\right]
 +
 \log\niab
 +
 \Suni{n}
\end{equation}
The first summand encodes now the detailed information about the family $i$, the second term indicates its \quotes{weight} in the entanglement spectrum \eqref{densityMatrix} and the third term,
\begin{align}\label{suni}
 \Suni{n}
 &:=
 \frac{1}{1-n}\log\left[\frac{\eta(q^n)}{\eta(q)^n}\right]\\
 &\approx 
 \frac{W}{12}\frac{n+1}{n}
 -
 \frac{1}{2}\log\left[\frac{W}{\pi}\right]
 +
 \frac{1}{2}\frac{\log n}{1-n}+\dots\notag
\end{align}
is universally appearing for all conformal families and counts the information contained in a Verma module. The leading order, obtained from $\eta\left(\tq^{1/n}\right)=\sqrt{n\pi/W}\,\eta(q^n)$, mimicks \eqref{SOmega}, aside from the effective central charge $\cce$.  Note also the appearance of a double logarithmic term $\log( W/\pi)$ prominent in $U(1)$ resolution \cite{Goldstein:2017bua}. The analysis here demonstrates firmly that the origin of this term is rooted in conformal symmetry. 

From \eqref{SRREnull} it is now clear that two distinct conformal families $i$ and $j$  can be equipartitioned for all $n$, i.e. $\Sab_n(i)=\Sab_n(j)$, if and only if
\begin{align}\label{ijEP}
 \niab=\nab^j 
 \quad
 \text{and}
 \quad
 \frac{\nul_i(q^n)}{(\nul_i(q))^n}=\frac{\nul_j(q^n)}{(\nul_j(q))^n}
\end{align}
This is called (exact) \emph{ij-equipartition}.
If this is the case for all families $i$ in the entanglement spectrum, $i\in\spec$, then the factorization $\iota_{\alpha\beta}$ is \emph{completely equipartitioned} \footnote{Exceptions are, of course, factorizations $\iota_{\ab}$ with only a single family, $\cH_{A,\ab}=\cH_i$. In which case the concept of equipartition is obsolete. One example of this is a $U(1)$-preserving Dirichlet-Dirichlet factorization $\iota_{DD}^{U(1)}$ in the free boson CFT at infinite radius \cite{DiGiulio:2022jjd}}. Exact equipartition is now analyzed in various CFTs:

%
a) Virasoro minimal models appear at $\cc(p,p')=1-6(p-p')^2/(pp')<1$ with $p,p'\in\Z_{\geq2}$ coprime. The families are labelled by two integers $1\leq r\leq p'-1$ and $1\leq s\leq p-1$ and have conformal weights $h_{(r,s)}=[(pr-p's)^2-(p-p')^2]/(4pp')$.
Each Verma module $\cV_{(r,s)}$ contains infinite null vectors leading to unique null structures $$\nul_{(r,s)}(q)=\sum_{k\in\Z}(q^{h_{(r,s+2pk)}}-q^{h_{(-r,s+2pk)}}).$$ 
Hence no two families can be exactly equipartitioned. 

b) Virasoro families at $\cc=1$: 
Verma modules are reducible at $\cc=1$ only for $h=h_m=m^2/4$ with $m\in\Z$, in which case $\nul_m(q)=q^{h_m}(1-q^{m+1})$. 
The vacuum, $h_{m=0}$, is of this type and has the null vector $L_{-1}\ket{0}$.
Two families $h_m,h_{m'}$ are never equipartitioned unless $m'=m$, as seen from \eqref{ijEP}. All other families, i.e. $h=h_\mu=\mu^2/4$ with $\mu\in\C\backslash\Z$, are non-singular. Thus they have $\nul_\mu(q)=q^{h_\mu}$ and are always equipartitioned, as long as their multiplicities in the entanglement spectrum coincide. Hence, an example of complete equipartition is easily found. Consider an XXZ spin chain. It corresponds to the free boson CFT at $\cc=1$ (Luttinger liquid) in the continuum. Picking a subinterval $A$ in the chain with free (N) boundary conditions at one end and fixed (D) boundary conditions at the other descends to a Neumann-Dirichlet (ND) factorization $\iota_{ND}$ in the CFT with entanglement spectrum \cite{Frohlich:1999ss}\footnote{$\iota_{ND}$ is the only factorization in the free boson which is independent of the compactification radius $R$ (Luttinger liquid parameter $K$).},
\begin{equation}\label{NDES}
 Z_{ND}(q)
 =
 \sum_{k=1}^\infty\chi_{h=\frac{(k-1/2)^2}{4}}(q)
 =
 \frac{1}{\eta(q)}\sum_{k=1}^\infty q^{\frac{(k-1/2)^2}{4}}
\end{equation}
It only contains Virasoro families of type $h_\mu$ with unit multiplicities. Therefore \eqref{SRREnull} assumes the same value for all families in $Z_{ND}$, $S^{ND}_n(h=\frac{(k-1/2)^2}{4})=\Suni{n}$. 

This result mimicks the well-known $U(1)$ resolution \cite{Goldstein:2017bua}. This is deceiving however, since both situations are entirely different on physical grounds. Indeed, $\Sab_n(i)=\Suni{n}$ can only be obtained in $U(1)$ resolution for $U(1)$-preserving factorizations $\iota_{\ab}$, which is not the case for \eqref{NDES} \cite{DiGiulio:2022jjd}. 
The lesson here is that Verma modules contain as much information as $U(1)$ familes, as should be since their characters are in fact identical. 

c) Unitary Verma factorizations: Define a \emph{Verma factorization} $\iota_{\ab}^\cV$ as one where all $i\in\spec$ have non-singular Verma modules $\cV_i$, i.e. $\nul_i(q)=q^{h_i}$. By eq. \eqref{SRREnull}, these are always equipartitioned once all multiplicities $\niab$ agree, in which case they can be normalized, $\niab=1$. The converse is not necessarily true once non-unitary families are involved \footnote{There can be completely equipartitioned non-Verma $\iota_{\ab}$. Imagine for instance a factorization at $0<\cc<1$ consisting of only two families with coinciding multiplicity and a null vector at the same level $N$, i.e. $\nul_i(q)=q^{h_i}(1-q^N)$. Such families satisfy \eqref{ijEP} and are non-unitary \cite{DiFrancesco:1997nk}.}. Thus, define \emph{unitary factorizations} $\iota_{\ab}^U$, as one for which all $i\in\spec$ are unitary families, i.e. $\cc=\cc(p,p')>0$ and $h_i=h_{(r,s)}$ or $\cc\geq1$ and $h_i\geq0$ \cite{DiFrancesco:1997nk}. Amongst the $\iota_{\ab}^{U}$, it is only the Verma factorizations, labelled $\iota_{\ab}^{U\cV}$, which are completely equipartitioned, once all $\niab=1$. This is gathered from the examples above and the following one.

d) Unitary factorizations at $\cc>1$:
The only unitary family with a null vector at $\cc>1$, namely $L_{-1}\ket{0}$, is the vacuum $\id$ \cite{DiFrancesco:1997nk}. This implies $\nul_\vac(q)=1-q$, while all other families are non-singular and have $\nul_{i\neq\vac}(q)=q^{h_i}$. A powerful consequence ensues. \emph{Any $\iota_{\ab}^{U}$ at $\cc>1$ with $\vac\notin\spec$ is Verma, i.e. $\iota_{\ab}^{U\cV}$, and thus completely equipartitioned so long as all multiplicities coincide}. A necessary condition for this is $\beta\neq\alpha$, which imposes boundary condition changing operators. 

From an information theoretic point of view, $\vir$ cannot discriminate amongst its unitary families with $h_i>0$ and $\cc>1$ via \eqref{SRE};
they store the same amount of information. This underpins the special role of the vacuum in gravity \cite{Hartman:2013mia, Faulkner:2013yia}, which is accessed by CFT in the holographic regime, $\cc\gg1$. 
Gravitational duals of Virasoro resolution must thus accentuate the vacuum.

\textit{7. Virasoro resolution of $\Sbdy^{\ab}$}.--- While the previous sections studied the SRRE \eqref{SRE} in isolation, the remainder of this letter investigates its consequences for the full entropy \eqref{Renyi}.

The von-Neumann entropy decomposes into the configuration $\Sconfig$ and fluctuation entropy $\Sfluct$ \cite{Lukin19,Wiseman03,Barghathi18,Barghathi:2019oxr,Kiefer-Emmanouilidis:2020otd,Kiefer-Emmanouilidis20Scipost,Monkman:2020ycn}, i.e. $S_1^{\alpha\beta}=\Sconfig^{\alpha\beta}+\Sfluct^{\alpha\beta}$, where
\begin{align}\label{EntanglementDecomposition}
 \Sconfig^{\alpha\beta}
 =
 \sum_{i\in\spec}\probab_iS^{\alpha\beta}_1(i)\,,
 \quad
 \Sfluct^{\alpha\beta}
 =
 -\sum_{i\in\spec}\probab_i\log\probab_i
\end{align}
While $\Sconfig$ collects and averages the entanglement stored within each family $i$, $\Sfluct$ accounts for the entanglement between the families. The relation $S_1^{\alpha\beta}=\Sconfig^{\alpha\beta}+\Sfluct^{\alpha\beta}$ is confirmed at all orders in the SM \cite{SM}. A profound information-theoretic relation is revealed when contemplating this decomposition asymptotically, i.e. in the limit $q\to1^-$, 
\begin{equation}\label{SbdyResolution}
 \Sbdy^{\ab}=\sum_{i\in\spec}\probab_i\left(\Stop^{\ab}(i)-\log\probab_i\right),
\end{equation}
where Eqs. \eqref{Renyi}, \eqref{SREasymp} and \eqref{probabilitesAsymp} have been employed. 
Indeed, the TEE \eqref{Stop} Virasoro-resolves the boundary entropy $\Sbdy^{\ab}$! Moreover, this result only requires the existence of a modular S-matrix, lifting \eqref{SbdyResolution} to a general lemma in rational CFTs, where $i$ labels families of an extended chiral symmetry algebra appearing in the entanglement spectrum.

\textit{8. Verma factorizations $\iota_{\ab}^{\cV}$}.---
To investigate complete equipartition further, it is useful to isolate the family-dependent terms in the entropies \eqref{Renyi} and \eqref{SRE} via $\sspe_n:=\Sab_n-\Suni{n}$ and $\sab_n(i):=\Sab_n(i)-\Suni{n}$.
This allows one to recast the entropies \eqref{EntanglementDecomposition} \cite{SM},
\begin{align}\label{ConfigFluct}
 \Sconfig^{\ab}
 &=
 \Suni{1}+\sum_{i\in\spec}\probab_i\sab_1(i)\\
 \Sfluct^{\ab}
 &=
 \sspe_1-\sum_{i\in\spec}\probab_i\sab_1(i)
\end{align}
Once $\sspe_1(i)=0$ for all $i\in\spec$, synonymous with $\iota_{\ab}^{\cV}$ and $\niab=1$, it follows that $\Sconfig^{\ab}=\Suni{1}$ and $\Sfluct^{\ab}=\sspe_1$. In this case, configuration and fluctuation entropy can be introduced at \emph{all} $n\in\N$, $\Sab_n=\Sconfign^{\ab}+\Sfluctn^{\ab}$ \cite{DiGiulio:2022jjd}, 
\begin{align}
 \Sconfign^{\ab}
 &:=
 \frac{1}{1-n}\log\tr[\rho_A(i)^n]
 =
 \Suni{n}\label{Sconfign}\\
 \Sfluctn^{\ab}
 &:=
 \frac{1}{1-n}\log\left[\sum_{i\in\spec}(\probab_i)^n\right]
 =
 \sspe_n\label{SFluctn}
\end{align}
These expressions are derived in the SM \cite{SM}.

As expected from a configuration entropy, eq. \eqref{Sconfign} accounts for the correlations within each family, resulting in the information contained in Verma modules. On the other hand, the entropy between the families $i$ is quantified by eq. \eqref{SFluctn}, which contains all information on the primaries of the theory. Clearly, $\Sfluctn^{\ab}=0$, if and only if $\spec$ consists of a single family. A glance at \eqref{ConfigFluct} reveals how this entanglement structure is deformed by familes with singular Verma modules. This introduces primary-dependence in $\Sconfign$ through null vector structures $\nul_i\neq q^{h_i}$.

\textit{9. Discussion and Outlook}.---
Many results of this letter, such as Eqs. \eqref{SRE}, \eqref{SREasymp}, \eqref{AsympEP}, \eqref{ijEP} or \eqref{SbdyResolution}, naturally lift to CFTs with extended symmetry, for instance Kac-Moody symmetry, by suitably replacing the modular S-matrix and characters. The prevalence of such CFTs \cite{DiFrancesco:1997nk,Gogolin:2004rp, Han:2019kyq, Recknagel:2013uja} enhances the applicability of the resolution discussed here tremendously. Equipartition is again analyzed by size comparisons of charge sectors. Note that the TEE appeared in (11) purely on grounds of conformal representations, and without relation to topological order. Hence, an intriguing continuation of the present work is to explore connections with such phases of matter. In these cases symmetry resolution should correspond to the extraction of entanglement in one anyon sector, promising insight into gapless \cite{feiguin2007interacting} and gapped topological systems \cite{cornfeld2019entanglement} including the quantum Hall effect \cite{protopopov2017transport}.

It will be interesting to apply the reasoning developed here to gravity. In general, it should be possible to reverse engineer constraints on the spectrum of a CFT from
its entanglement spectrum. Unfortunately, Virasoro resolution can only distinguish the vacuum, so it is worthwhile to generalize the present construction to extended symmetries attuned to gravity, such as higher spin symmetry \cite{gaberdiel2013minimal}. The study of its entanglement spectrum should provide constraints on the spectrum of $(2+1)d$ gravity \cite{Witten:2021nzp,Basile:2023ycy}.

Finally, using the g-theorem \cite{Friedan:2003yc} it will be interesting to study whether \eqref{SbdyResolution} has implications for $\Stop^{\ab}(i)$ under boundary renormalization group flows.
\bigskip

\textbf{Acknowledgements}.--- It is a pleasure to thank Ilka Brunner, Ram Brustein, Shira Chapman, Saskia Demulder, Giuseppe di Giulio, Ren\'e Meyer, Ingo Runkel, Henri Scheppach, Suting Zhao for fruitful discussions. Moreover, I am grateful to Saskia Demulder, Giuseppe di Giulio and Suting Zhao for a careful reading of the draft. My work is supported by the Israel Science Foundation (grant No. 1417/21) and by the German Research Foundation through a German-Israeli Project Cooperation (DIP) grant “Holography and the Swampland” and by Carole and Marcus Weinstein through the BGU Presidential Faculty Recruitment Fund.


\begin{widetext}
\appendix
\section{Characters and null vectors}\label{appCharacters}
The most important object in the study presented in the main text is a conformal character
\begin{align}\label{character}
 \chi_i(q)
 &=
 \tr_{i}\left[q^{L_0-\frac{\cc}{24}}\right]
 =
 q^{h_i-\cc/24}\sum_{m}d(m)q^m,
\end{align}
where $\tr_i:=\tr_{\cH_i}$ and $\cH_i:= \cH(\cc,h_i)$ is an irreducible Virasoro modules at conformal weight $h_{i}$ and central charge $\cc$. A character can be seen as \quotes{mini} partition function, where the state space is only $\cH_i$. The functions $d(m)$ are integer degeneracy factors collecting the number of states at fixed energy $m$. 

In the simplest case the degeneracies $d(m)$ are the partitions $p(m)$ of the integer $m$. These state spaces are called Verma modules $\cV_i$ and are constructed as follows. Starting from a primary $\ket{h_i}$, a Verma module $\cV_i$ is built from all linear independent descendants $L_{-k_1}L_{-k_2}\dots L_{-k_l}\ket{h_i}$. The character \eqref{character} of a Verma module $\cV_i$ can be evaluated, 
\begin{equation}
\chi_i^{\verm}=\frac{q^{h_i+(1-\cc)/24}}{\eta(q)}
\end{equation}
where $\eta(q)=\prod_{k=1}^\infty(1-q^k)$ is the Dedekind eta function, which counts partitions $p(m)$ at all energy levels $m$. 

In many relevant physical applications, it may happen that a specific descendant $\xi$ at level $N$ is simultaneously primary. This is called a null vector and it furnishes its own Verma module $\cV_\xi$, which stands orthogonal to all other states generated from $\ket{h_i}$. Therefore it decouples from $\cV_i$ and can be quotiented out. An irreducible Virasoro module $\cH_i$ is obtained after appropriately quotienting out all null vectors from $\cV_i$. Clearly, this proceedure reduces the size of the vector space, and thus $d(m)\leq p(m)$ in \eqref{character}.

This is reflected in the characters of the irreducible module $\cH_i$. Consider for instance the case of a single null vector $\xi$ at level $N$, which has been quotiented out. Note that the null field $\xi$ has conformal weight $h_\xi=h_i+N$. The original Verma module $\cV_i$ is \quotes{rid} of the Verma module $\cV_\xi$,
\begin{equation}
\chi_i
=
\chi_i^{\verm}-\chi_\xi^\verm
=
\frac{q^{h_i+\frac{1-\cc}{24}}}{\eta(q)}-\frac{q^{h_\xi+\frac{1-\cc}{24}}}{\eta(q)}
=
\frac{q^{h_i+\frac{1-\cc}{24}}}{\eta(q)}(1-q^N)
\end{equation}
When more than one singular vector is present, care must be taken, since the submodules of the null vectors can overlap, as is the case for minimal models. A thorough analysis is found in chapter 8 of \cite{DiFrancesco:1997nk}.

For the purposes of the main text, the character for $\cH_i$ is rewritten in terms of a function $\nul_i(q)$ which encodes the null vector structure
\begin{equation}\label{nullStructureApp}
 \chi_i(q)=\frac{q^{\frac{1-\cc}{24}}}{\eta(q)}\nul_i(q)
\end{equation}
For instance, when $\cH_i$ is a Verma module, i.e. it has no singular vectors, then $\nul_i(q)=q^{h_i}$, and when $\cH_i$ carries a single null vector at level $N$, $\nul_i(q)=q^{h_i}(1-q^N)$.

\section{Quantum dimensions}\label{appQdim}
The quantum dimensions
\begin{equation}\label{qdimApp}
\qdim_i=\lim_{q\to1^-}\frac{\chi_i(q)}{\chi_{\vac}(q)}=\frac{\modS_{i\Omega}}{\modS_{\vac\Omega}} 
\end{equation}
measure the \quotes{asymptotic size} of the module $i$ in relation to the vacuum module. Use is made of the modular S-transformation defined by
\begin{equation}\label{SmatrixApp}
 \chi_i(q^n)
 =
 \sum_{k\in I}(\modS^{-1})_{ik}\chi_k(\tq^{1/n})
 \approx
 \modS_{i\Omega}\, \chi_{\Omega}(\tq^{1/n}),
 \qquad
 (\modS^{-1})_{i\Omega}
 =
 (\modS^{*})_{i\Omega}
 =
 \modS_{i\Omega}
 =
 \modS_{\Omega i}>0.
\end{equation}
The approximation uses the fact that the ground state character dominates the sum in the limit $\tq\to0$.
Note that this expression contains a linear sum over families $i\in I$ ($I$ is the set of all conformal families at fixed central charge $\cc$), and hence the modular S-matrix and the quantum dimensions are useful objects in rational CFTs, where this sum is finite. The total quantum dimension is defined by $\qdim^2=\sum_{i\in I}\qdim_i^2$. Because $\modS_{i\Omega}=\modS_{\Omega i}>0$, it follows that $\qdim_i>0$. For theories with self-conjugate representations only -- as is the case for Virasoro -- it can be shown that $\modS_{\vac\Omega}=1/\qdim$ and thus $\modS_{i\Omega}=\qdim_i/\qdim$. 

Indeed, the modular S-matrix squares to the charge conjugation matrix, $\modS^2=\mathsf{C}$, or equivalently $\sum_{k\in I}\modS_{ik}\modS_{kj}=\delta_{j,i^+}$. $\vir$ only has selfconjugate representations, $i^+=i$. It follows that $\sum_{k\in I}\modS_{\Omega k}\modS_{k\Omega}=\delta_{\Omega\Omega}=1$. This yields 
\begin{equation}
 \qdim^2=\frac{1}{(\modS_{\vac\Omega})^2}\sum_{k\in I}(\modS_{k\Omega})^2=\frac{1}{(\modS_{\vac\Omega})^2}
\end{equation}
and establishes the first claim.  The remaining claim, $\modS_{i\Omega}=\qdim_i/\qdim$, follows immediately from \eqref{qdimApp}. For unitary theories $\Omega=\id$, so that \eqref{qdimApp} reduces to the standard definition in the literature.

\section{Asymptotic entropy expansions}\label{appAsympEntropy}
Key to deriving all asymptotic expressions in the main text, i.e. those for which $\tq\to0^+$ or equivalently $W\to\infty$, $\epsilon\to0^+$ is the asymptotic behavior of a character \eqref{SmatrixApp}. It implies for the parition function of the entanglement spectrum of the factorization $\iota_{\ab}$ that
\begin{equation}\label{SpectrumApproxApp}
 \Zab(q^n)
 =
 \sum_{i\in\spec}\niab\chi_i(q^n)
 =
 \sum_{i\in\spec}\sum_{k\in I}\niab(\modS^{-1})_{ik}\chi_k(\tq^{1/n})
 \approx
 \left(\sum_{i\in\spec}\niab\modS_{i\Omega}\right)\chi_\Omega(\tq^{1/n})
\end{equation}
The term in parenthesis is in fact the Affleck-Ludwig g-factor. To see this, boundary states $\bket{\alpha}=\sum_{i\in\tilde{\spec}}\cB_\alpha^i\iket{i}$ have to be recalled, where $\tilde{\spec}$ is a set of families containing $\Omega$ \footnote{There is an implicit requirement imposed in this argument, namely that the ground state $\Omega$ has $h_\Omega=\bh_\Omega$. For the unitary theories, $\Omega=\id$ and this is automatically satisfied.}. They store the full information of the boundary condition $\alpha$ in terms of bulk states represented by Ishibashi states $\iket{i}$. The details are not important, but interested readers are referred to references on boundary conformal field theory, e.g. \cite{Recknagel:2013uja}. What matters is that the coefficient $\cB_\alpha^\Omega$ for the ground state $\Omega$ is the g-factor $\gf_\alpha$ for the boundary condition $\alpha$. It roughly counts the degrees of freedom living on the boundary $\alpha$.

The Cardy constraint \cite{Cardy:1989ir},
\begin{equation}
 \sum_{j\in\tilde{\spec}}\cB^j_\alpha\cB^j_\beta\chi_j(\tq)
 =
 \sum_{i\in \spec}\sum_{j\in I}\niab (\modS^{-1})_{ij}\chi_j(\tq)
\end{equation}
can be evaluated asymptotically and immediately yields $\gf_\alpha\gf_\beta=\sum_{i\in\spec}\niab\modS_{i\Omega}$ and hence \eqref{SpectrumApproxApp} becomes $\Zab(q^n)\approx\gf_\alpha\gf_\beta\,\chi_\Omega(\tq^{1/n})$. This is used to derive the asymptotic expression of the R\'enyi entropies in the main text,
\begin{equation}
 \Sab_n
 =
 \frac{1}{1-n}\log\left[\frac{\Zab(q^n)}{(\Zab(q))^n}\right]
 \approx
 \log[\gf_\alpha\gf_\beta]
 +
 \frac{1}{1-n}\log\left[\frac{\chi_\Omega(\tq^{1/n})}{(\chi_\Omega(\tq))^n}\right]+\dots
 =
 \Sbdy^{\ab}+S_n^{\Omega}+\dots
 \label{RenyiApp}
\end{equation}
This analysis is simpler for the SRRE since the asymptotic expansion in \eqref{SmatrixApp} may directly be employed,
\begin{equation}
 \Sab_n(i)
 =
 \frac{1}{1-n}\log\left[\frac{\chi_i(q^n)}{(\chi_i(q))^n}\right]
 \approx
 \log[\modS_{i\Omega}\niab]
 +
 \frac{1}{1-n}\log\left[\frac{\chi_\Omega(\tq^{1/n})}{(\chi_\Omega(\tq))^n}\right]+\dots 
 =
 \Stop^{\ab}(i)+S_n^{\Omega}+\dots
 \label{SRREapp}
\end{equation}

\section{Entropy Decompositions}\label{appEntropyDecomp}
In this section, the entropy decomposition into configuration and fluctuation entropy,
\begin{align}
 S_1^{\alpha\beta}
 =
 \Sconfig^{\alpha\beta}+\Sfluct^{\alpha\beta}\,,
 \quad
 \Sconfig^{\alpha\beta}
 =
 \sum_{i\in\spec}\probab_i\Sab_1(i),
 \quad
 \Sfluct^{\ab}
 =
 -\sum_{i\in\spec}\probab_i\log\probab_i
 \label{EntanglementDecompositionApp}
\end{align}
is checked to all orders and asymptotically, beginning with the latter.

First of all, the probabilities can be directly approximated through use of \eqref{SmatrixApp} and \eqref{SpectrumApproxApp} for $n=1$ and expressed in various ways,

\begin{equation}\label{probabilitiesApp}
 \probab_i
 =
 \tr[\Pi_i\rho_A]
 =
 \niab\frac{\chi_i(q)}{Z_{\alpha\beta}(q)}
 \approx
 \frac{\niab\modS_{i\Omega}}{\sum_{j\in\spec}\nab^j\modS_{j\Omega}}
 =
 \frac{\niab\modS_{i\Omega}}{\gf_\alpha\gf_\beta}
 =
 \frac{\niab\qdim_{i}}{\sum_{j\in\spec}\nab^j\qdim_j},
\end{equation}
Reassuringly, they still satisfy $\sum_{i\in\spec}\probab_i=1$. Hence, after plugging the right hand sides of \eqref{RenyiApp} and \eqref{SRREapp} for $n=1$, respectively, into \eqref{EntanglementDecompositionApp}, $S_1^\Omega$ cancels out and 
\begin{align}
 \Sbdy^{\alpha\beta}
 =
 \sum_{i\in\spec}\probab_i(\Stop^{\ab}(i)-\log\probab_i),
 \label{AsympEntanglementDecompositionApp}
\end{align}
is revealed. This is the Virasoro resolution of boundary entropy via topological entanglement entropies highlighted in the main text.  Note that this derivation did not require explicit knowledge of $S^\Omega_1=\lim_{n\to1}S^\Omega_n$.

In order to confirm \eqref{EntanglementDecompositionApp} to all orders it is convenient to observe that the probabilities provide an alternate expression for the partition function
\begin{equation}\label{Auxiliary}
 \frac{\niab\chi_i(q)}{\probab_i}
 =
 Z_{\alpha\beta}(q)
 =
 \frac{\nab^j\chi_j(q)}{\probab_j}
\end{equation}
The LHS and RHS of this equation are thus constants amongst conformal families. Another ingredient is, of course, the explicit form of the symmetry-resolved von-Neumann entropy
\begin{equation}
 \Sab_1(i)
=
\lim_{n\to1}\frac{1}{1-n}\log\left((\niab)^{1-n}\frac{\chi_i(q^n)}{(\chi_i(q))^n}\right)
=
\log[\niab\chi_i(q)]-q\log(q)\frac{\chi_i'(q)}{\chi_i(q)}\label{SRE1App}
\end{equation}
where the prime indicates a derivative with respect to $q$. This expression needs to be recovered in the (full) von-Neumann entropy
\begin{align}
 \Sab_1
 &=
 \lim_{n\to 1}\frac{1}{1-n}\log\left[\frac{\Zab(q^n)}{(\Zab(q))^n}\right]\notag\\
 &=
 -q\log(q)\frac{\Zab'(q)}{\Zab(q)}+\log\Zab(q)\notag\\
 &=
 -q\log(q)\sum_{i\in\spec}\frac{\niab\chi'_i(q)}{\Zab(q)}-\sum_{i\in\spec}\probab_i\log\Zab(q)\notag\\
 &=
 -q\log(q)\sum_{i\in\spec}\probab_i\frac{\chi'_i(q)}{\chi_i(q)}-\sum_{i\in\spec}\probab_i\log\left[\frac{\niab\chi_i(q)}{\probab_i}\right]\notag\\
 &=
 \sum_{i\in\spec}\probab_i\left(\Sab_1(i)-\log\probab_i\right)\label{ProofDecomposition}
\end{align}
The second line evaluates the limit of the first line, the third line uses that the definition of $\Zab$ and that probabilities sum to unity. The fourth line uses \eqref{Auxiliary}; in particular the fact that the partition function can be expressed as said ratio for different families $i$. The last line reorganizes terms such that \eqref{SRE1App} is recovered, thereby confirming \eqref{EntanglementDecompositionApp}.

\section{(Symmetry-resolved) R\'enyi entropies in terms of $\Suni{1}$ and $\sspe_1$}
Recall the definitions of $\sspe_n$ and $\sspe_n(i)$, which are presented here as rewritings of the (symmetry-resolved) R\'enyi entropies
\begin{align}
 \Sab_n
 &=
 \Suni{n}+\sspe_n
 =
 \Suni{n}
 +
 \frac{1}{1-n}\log\left[\frac{\sum_{i\in\spec}\niab\nul_i(q^n)}{(\sum_{j\in\spec}\nab^j\nul_j(q))^n}\right]\label{RenyiChargeOnlyApp}\\
 \Sab_n(i)
 &=
 \Suni{n}+\sab_n(i)
 =
 \Suni{n}
 +
 \frac{1}{1-n}\log\left[\frac{\nul_i(q^n)}{\nul_i(q)^n}\right]
 +
 \log\niab\label{SREchargeOnlyApp}
\end{align}
The definition of $\Sconfig^{\ab}$ in \eqref{EntanglementDecompositionApp} readily yields
\begin{align}
 \Sconfig^{\ab}
 =
 \sum_{i\in\spec}\probab_i\Sab_1(i)
 =
 \Suni{1}+\sum_{i\in\spec}\probab_i\sspe_1(i)
\end{align}
To re-express the fluctuation entropy first note that \eqref{ProofDecomposition} becomes
\begin{align}
 \sspe_1
 =
 \sum_{i\in\spec}\probab_i(\sspe_1(i)-\log\probab_i)
 =
 \sum_{i\in\spec}\probab_i\sspe_1(i)+\Sfluct^{\ab}
\end{align}
which becomes $\Sfluct^{\ab}=\sspe_1-\sum_{i\in\spec}\probab_i\sspe_1(i)$ as claimed in the main text.

\section{Configuration and fluctuation entropy at all $n$ for complete equipartition}

Once $\sspe_1(i)=0$ for all $i\in\spec$, synonymous with $\iota_{\ab}^{\cV}$ and $\niab=1$, it follows that $\Sconfig^{\ab}=\Suni{1}$ and $\Sfluct^{\ab}=\sspe_1$. In this case, configuration and fluctuation entropy can be introduced at \emph{all} $n\in\N$, $\Sab_n=\Sconfign^{\ab}+\Sfluctn^{\ab}$ \cite{DiGiulio:2022jjd}, 
\begin{align}
 \Sconfign^{\ab}
 &:=
 \frac{1}{1-n}\log\tr[\rho_A(i)^n]
 =
 \Suni{n}\label{Sconfign}\\
 \Sfluctn^{\ab}
 &:=
 \frac{1}{1-n}\log\left[\sum_{i\in\spec}(\probab_i)^n\right]
 =
 \sspe_n\label{SFluctn}
\end{align}
To reach the first line, Eq. (10) of the main text is supplemented by $\sspe_n(i)=0$. The second line employs $\nul_i=q^{h_i}$ in \eqref{RenyiChargeOnlyApp} and that the probabilities $\probab_i=\niab\chi_i(q)/Z_{\alpha\beta}(q)$ are simplified by use of \eqref{nullStructureApp} to $\probab_i=\niab\nul_i(q)/(\sum_{k\in\spec}\nab^k\nul_k(q))$. 
\section{Asymptotic equipartition in Virasoro minimal models}
The only rational CFTs with symmetry $\vir\times\vir$ are the Virasoro minimal models \cite{Belavin:1984vu} appearing at central charge $$\cc(p,p')=1-6\frac{(p-p')^2}{pp'}$$ with $p,p'\in\Z_{\geq2}$ coprime. The families are labelled by two integers $1\leq r\leq p'-1$ and $1\leq s\leq p-1$ and have conformal weights $$h_{(r,s)}=\frac{(pr-p's)^2-(p-p')^2}{4pp'}.$$ This enjoys the symmetry $h_{(p'-r,p-s)}=h_{(r,s)}$ so that there are $(p'-1)(p-1)/2$ independent fields. The ground state $\Omega=(r_\Omega,s_\Omega)$ is the one with lowest conformal weight. For unitary theories this is always $\id=(1,1)$. Each Verma module associated with these families contains infinite null vectors, resulting in the structure $$\nul_{(r,s)}(q)=\sum_{k\in\Z}(q^{h_{(r,s+2pk)}}-q^{h_{(-r,s+2pk)}}).$$ Eq. (18) of the main text thus negates exact equipartition of any two families.

Asymptotic equipartition can occur, however, for families $i=(r,s)$ and $j=(r',s')$. Since the remainder of this section deals with asymptotic equipartition, the condition for its fulfillment is repeated here for convenience,
\begin{equation}\label{AsympEPApp}
 \Stop^{\ab}(i)=\Stop^{\ab}(j)
 \qquad 
 \Leftrightarrow
 \qquad
 \qdim_i=\qdim_j
 \text{ and }
 \niab=\nab^j
\end{equation}
for $i,j\in\spec$ and $i\neq j\,$. In the main text, this is Eq. (13).

The quantum dimensions $\qdim_i=\modS_{i\Omega}/\modS_{\Omega\Omega}$ are analyzed with the modular S-matrix
\begin{equation}
 \modS_{(r_1,s_1),(r_2,s_2)}
 =
 -\sqrt{\frac{8}{pp'}}(-)^{r_1s_2+s_1r_2}
 \sin\left(\frac{p}{p'}r_1r_2\pi\right)
 \sin\left(\frac{p'}{p}s_1s_2\pi\right)
\end{equation}
This is expression lies always in $(-1,1)$ except for the trivial theory $\cc(p=3,p'=2)=0$, which has only the identity field $(1,1)=\id$ and $\modS_{\id\id}=1$. This case is excluded in the following. Together with $\nab^{(r,s)}\in\N$, asymptotic equipartition \eqref{AsympEPApp} then demands that the multiplicities $\nab^{(r,s)}$ and quantum dimensions $\qdim_{(r,s)}$ agree respectively for families $(r,s),(r',s')$.

It is readily checked that \footnote{These solutions are not exhaustive.} the families $(r,s),\, (t'-r,s)$ and $(r,t-s)$ have equal quantum dimensions, i.e. $\modS_{(r,s),\Omega}=\modS_{(t'-r,s),\Omega}=\modS_{(r,t-s),\Omega}$, if
\begin{align}\label{ts}
 \Z\ni t'
 =
 \frac{p'(1+2c_1)}{\lambda_\Omega},
 \qquad
 \Z\ni t
 =
 \frac{p(1+2c_2)}{\lambda_\Omega},
 \qquad
 \lambda_\Omega=p\,r_\Omega+p's_\Omega\,.
\end{align}
Constants $c_{1,2}\in\Z$ need to be found such that $1+r\leq t'\leq p'+r-1$ and $1+s\leq t\leq p+s-1$. When $\lambda_\Omega$ is odd, $t'=p'$ and $t=p$ may be chosen. For the fields $i=(r,s)$ and $j=(t'-r,s)$ to truly be distinct, $t'\neq2r$ has to be excluded and when $2s=p$ one must also exclude $t'=p'$. For the fields $i=(r,s)$ and $j=(r,t-s)$ to truly be distinct, $t\neq2s$ has to be excluded and when $2r=p'$ one must also exclude $t=p$. Finally, the fields $(t'-r,s)$ and $(r,t-s)$ are identical if either $t=p$ and $t'=p'$ or $t'=2r$ and $t=2s$. 

The key to finding these solutions is to realize that, e.g. $\modS_{(r,s),\Omega}=\modS_{(r',s),\Omega}$ reduces to 
\begin{equation}\label{EqualQdims}
 \sin\left(\frac{\pi r}{\lambda_\Omega}\right)=\sin\left(\frac{\pi r'}{\lambda_\Omega}\right)
\end{equation}
after applying $(-1)^k\sin(x)=\sin(x+\pi k)$ for $k\in\Z$. Since $r,r'$ are integer, $r'=r+\ell$ for $\ell\in\Z$. Plugging this into \eqref{EqualQdims} and looking for non-trivial solutions, provides $\ell=t'-2r$ with $t'$ as in \eqref{ts}, leading to $r'=t'-r$.

To investigate the multiplicities $\nab^{(r,s)}$, the entanglement spectrum or equivalently the factorization must be chosen. I choose a factorization made up of Cardy states and call that a \emph{Cardy factorization} $\iota_{\alpha\beta}^{\mathscr{C}}$. This requires a diagonal bulk modular invariant, i.e. $\Zbulk(q,\bq)=\sum_{i\in I}|\chi_i(q)|^2$ \footnote{When non-charge-conjugate sectors are present, a charge conjugate modular invariant may also be chosen, $M_{i\bi}=\delta_{\bi,i^+}$. For $\vir$ one always has $i^+=i$.},  For these, the boundary conditions are labelled by the primaries $I\ni\alpha=(r_\alpha,s_\alpha)$ of the theory and the multiplicities are the fusion coefficients \cite{Cardy:1989ir},
\begin{align}\label{fusionRules}
 \mathsf{n}_{(r_\alpha,s_\alpha)(r_\beta,s_\beta)}^{(r,s)}
 =
 \fus_{(r_\alpha,s_\alpha)(r_\beta,s_\beta)}^{(r,s)}
 =
 \begin{cases}
  1 &\text{ if } |r_\alpha-r_\beta|\leq r\leq \min(r_\alpha+r_\beta,2p'-r_\alpha-r_\beta)
  \\
   &\text{ \& } |s_\alpha-s_\beta|\leq s\leq \min(s_\alpha+s_\beta,2p-s_\alpha-s_\beta)
   \\
   0 &\text{ otherwise. }
 \end{cases}
\end{align}
where $r+r_\alpha+r_\beta$ and $s+s_\alpha+s_\beta$ have to be odd. 

The inequalities on the RHS of \eqref{fusionRules} are to be read as constraints on the factorization, i.e. on $r_{\alpha,\beta}, s_{\alpha,\beta}$, after having found pairs of families $(r,s)\,\&\,(t'-r,s)$ or $(r,s)\,\&\,(r,t-s)$ with coinciding quantum dimension. In the former case, familes $\alpha,\,\beta$ need to be found, which contain $(r,s)\,\&\,(t'-r,s)$ in their fusion. In the latter case, familes $\alpha,\,\beta$ need to be found, which contain $(r,s)\,\&\,(t'-r,s)$ in their fusion.

To get a feeling for these solutions, restrict to unitary minimal models with $p=p'+1=m+1$. In this case, $\Omega=\id=(1,1)$ and thus $\lambda_\Omega=p+p'=2m+1$. Choosing $c_1=c_2=m$ for simplicity leads to $t'=p'$ and $t=p$ in \eqref{ts}. The remaining two fields $(t'-r,s)=(p'-r,s)$ and $(r,t-s)=(r,p-s)$ are in fact the same field, by standard identification on the Kac table. Hence, it suffices to restrict to analyzing the former. By virtue of the discussion above, the two families $(r,s)$ and $(p'-r,s)\simeq(r,p-s)$ have equal quantum dimensions, $\modS_{(r,s),\id}=\modS_{(p'-r,s),\id}$. For these to truly be distinct familes, the cases $2r=p'=m$ for even $m$ and $2s=p=m+1$ for odd $m$ have to be excluded. Cardy Factorizations, i.e. the ones with \eqref{fusionRules}, must thus have $\ab$ satisfying $\fusab^{(r,s)}=\fusab^{(p'-r,s)}=\fusab^{(r,p-s)}=1$

For example, consider a Cardy factorization for the Ising CFT, which is a unitary minimal model with $p=4,p'=3$. It has three primaries $\id=(1,1)$, $\sigma=(2,2)$ (not to be confused with the label for the entanglement spectrum) and $\varepsilon=(2,1)$. Hence, \eqref{AsympEPApp} checks for the S-matrix elements $\modS_{\id,(r,s)}$. The class of solutions described in the previous paragraph selects $r=1,2$ and $s=1,3$ so that $\modS_{\id\id}=\modS_{\id\varepsilon}=1/2$. The non-trivial fusion rules are $\varepsilon\star\varepsilon=\id$, $\varepsilon\star\sigma=\sigma$ and $\sigma\star\sigma=\id+\varepsilon$. The latter shows that the factorization $\iota_{\sigma\sigma}^{\mathscr{C}}$ with entanglement spectrum $Z_{\sigma\sigma}=\chi_\id+\chi_\varepsilon$ features asymptotic $\id\varepsilon$-equipartition.
\end{widetext}

\bibliographystyle{apsrev4-2}
\bibliography{Bibliography.bib}

\end{document}